\documentclass{article}
\usepackage[latin1]{inputenc}  
\usepackage{focus}
\usepackage{graphicx} 
\usepackage{mdwlist} 
\usepackage[usenames,dvipsnames]{xcolor}

\newcommand{\amessage}{AMessage}
\newcommand{\cmessage}{Message}

\newcommand{\focust}{\textsc{Focus}$^{ST}$}

\title{Formal \focust\ Specification of CAN}
\author{Maria Spichkova}

\begin{document}

\maketitle

\begin{abstract}
This paper presents a formal specification of the Controller Area Network (CAN) protocol using \focust\ framework.   
We formally describe core components of the protocol, which provides a basis for further formal analysis using the Isabelle/HOL theorem prover.
\end{abstract}

\section{Introduction}
\label{section:intro}

Controller Area Network (CAN) protocol is one of the standard communication protocols used in automotive systems. CAN was developed by Robert Bosch GmbH \cite{bosch1991can} and is a part of the ISO 11898 standard \cite{organizacion1993iso}.

In this paper, we present a formal specification of this protocol using \focust\ framework.   
 \focust\  was introduced as an extension of the \Focus\ language, see \cite{focus,spichkova2014modeling}.   
Similarly to \Focus, specifications in \focust\ are based on the notion of \emph{streams}, and a formal meaning of a specification is exactly this external \emph{input/output relation}.
However, in the original \Focus  input and output streams of a component are mappings
 of natural numbers  to single messages,
whereas a \focust\  stream %
 is a mapping from natural numbers  to lists of messages  within the corresponding time intervals.
 Moreover, the syntax of \focust\ is particularly devoted to specify spatial (S) and timing (T) aspects in a comprehensible fashion,
which is the reason to extend the name of the language by $^{ST}$.

The \focust\ specification layout was then discussed  in  \cite{spichkova2016spatio}. Here, we present only a small subset of that we applied to specify the CAN protocol:
\begin{itemize}

\item $\nempty$ denotes an empty stream;

\item $\ndom{s}$ yields the list $[1...\#s]$, where $\#s$ denotes the length of the stream $s$; 

\item $\nrng{s}$ converts the stream $s$ into a set of its elements : $\{s.j | j \in \ndom{s}\}$;

\item 
The predicate $\msg{n}{s}$ is $\ntrue$ iff the stream $s$ has at every time interval 
at most $n$ messages.
\end{itemize} 

 \begin{figure}[ht!]
    \centering
    \begin{spec}{\spc{SystemArch}}{gb}
    \includegraphics[scale = 0.7]{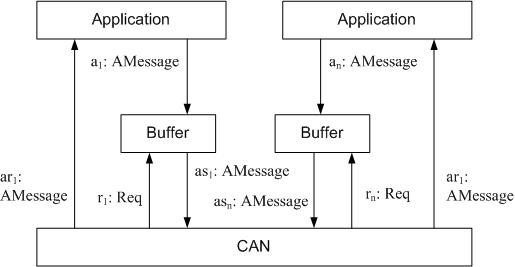}
    \end{spec}
    \caption{Logical architecture of a CAN-based system}
    \label{fig:CANarch}
\end{figure}

\section{Specification of a CAN-based system}

Figure \ref{fig:CANarch} presents the specification $SystemArch$, which describes a logical architecture of a CAN-based system. We define the following the data types for this specification:
$\amessage$ represents the data type of messages, which are sent by one automotive application to another:
\[ 
	\ntype \amessage = msg(id:~\Nat,~data:~Data)
\]
$\cmessage$ will denote the CAN-internal messages, and $Req$ will be a simple request type to denote the CAN requsts to the system bufferes.
\[
	\ntype \cmessage = \Nat \mid Data
\]
\[
	\ntype Req = \Nat
\]
 
The core system requirements are defined by the following specification $CAN$, where the assumption is that all data streams $as_i$ (which CAN receives from the automotive application components via the corresponding  buffer components) satisfy the $\msg{1}{as_i}$ predicate, i.e., all these streams must have at every time interval 
at most one message. The guarantee part of this specification has two predicates that define
\begin{enumerate}
    \item 
    all data streams $ar_i$ (which CAN sends to the the corresponding automotive application components) satisfy the $\msg{1}{ar_i}$ predicate,
    \item 
    the data transmission is correct as per the predicate \textit{MessageTransmission}. 
\end{enumerate} 

\begin{spec}{\spc{CAN}}{td}
\InOut
{as_1, \dots,  as_n: \amessage}
{ar_1, \dots,  ar_n: \amessage;~r_1, \dots,  r_n: Req}
\uasm 
\\ 
\abox{1}  \t1  \forall i \in [1..n].~\msg{1}{as_1}
\\
\zeddashline 
\ugar \\
\cbox{1}  \t1  \forall i \in [1..n].~\msg{1}{ar_1}
 \\
\cbox{2}  \t1 MessageTransmission(as_1, \dots,  as_n, ar_1, \dots,  ar_n, r_1, \dots,  r_n)
\end{spec}

\noindent
Note, that in contrast to the \focust specification of FlexRay  \cite{kuhnel2006flexray,kuhnel2006upcoming,kuhnel2007fault,spichkova2017formal, spichkova2018focusst}, where
the correct transmission means the transmission according the FlexRay 
scheduling tables, in the case of CAN the correct transmission is specified according the priority relations, see below.

\begin{schema}{\spc{MessageTransmission}}
as_1, \dots,  as_n: \nist{\amessage};\\
ar_1, \dots,  ar_n: \nist{\amessage};~r_1, \dots,  r_n: \nist{Req}
\ST
\forall t \in \Nat:\\
~\\
\auxbox{1}  \t1   (\forall i \in [1..n]: \ti{as_i}{t} = \nempty) ~\to~
    \forall j \in [1..n]: \ti{ar_j}{t+2} = \nempty
\\
~\\
\auxbox{2}  \t1  \forall i, j \in [1..n]: \ti{ar_i}{t} = \ti{ar_j}{t}
\\
~\\
\auxbox{3}  \t1   \exists i \in [1..n]: \ti{as_i}{t} \neq \nempty \wedge id(\nft{\ti{as_i}{t}}) = MinNatList(x) \to\\
    \t2 \ti{r_i}{t+2} \neq \nempty ~\wedge~ \forall j \in [1..n]: \ti{ar_j}{t+2} = \ti{as_i}{t} 
\\
~\\
\nwhere ~x = TakeIds(y)\\
\t3 y = CollectElements(n,\ti{as_1}{t}, \dots, \ti{as_n}{t})\\
\end{schema}

We also defined the following auxiliary functions to specify the $MessageTransmission$ predicate:
\begin{itemize}
    \item 
    $TakeIds$ takes as an input a finite list of type $\amessage$ and returns the corresponding finite list of the identifiers.
    \item
    $CollectElements$ describes collection of all data received by CAN at a particular time interval.
    \item 
    $MinNatList$ finds the smallest element in a finite list of natural numbers.
\end{itemize}

\begin{schema}{\spc{TakeIds}}
\nfst{\amessage} \to \nfst{\Nat} \\
\ST
\auxbox{1}  \t1 TakeIds(\nempty) = \nempty
\\
\auxbox{2}  \t1 TakeIds(\angles{x} \nconc y) = \angles{id(x)} \nconc TakeIds(y) 
\end{schema}

\begin{schema}{\spc{CollectElements}}
\Nat \times \dots \times \nfst{M} \times \nfst{M} \to \nfst{M} \\
\ST
\auxbox{1}  \t1 CollectElements(0, s_1,\dots, s_n) = \nempty
\\
\auxbox{2}  \t1 CollectElements(i+1, s_1,\dots, s_n) =\\
\t3 \nif s_{i+1} = \nempty\\
\t3 \nthen CollectElements(i, s_1,\dots, s_n)\\
\t3 \nelse s_{i+1} \nconc CollectElements(i, s_1,\dots, s_n)\\
\t3 \nfi
\end{schema}

\begin{schema}{\spc{MinNatList}}
\Nat \times \dots \times \nfst{\Nat} \to \Nat \\
\ST
\auxbox{1}  \t1 MinNatList(a, \nempty) = a
\\
\auxbox{2}  \t1 MinNatList(a, \angles{x}\nconc y) = \\
\t3 \nif a \le x \\
\t3 \nthen MinNatList(a, y)\\
\t3 \nelse MinNatList(x, y)\\
\t3 \nfi
\end{schema}

We specify a CAN-buffer in \focust\ as a component $Buffer$, see below.
This component has two input streams (data from an automotive application and requests from CAN). The only assumption on the inputs is that the data stream from an automotive application must have at most one message per each time unit. The output stream will also have at most one message per each time unit.  In the even time intervals, the buffer's output stream will be empty, where in the even time intervals it will send the stored data to the CAN component.

\newpage 

\begin{spec}{\spc{Buffer}}{td}
\InOut{a~\amessage;~r: Req}{as: \amessage}
\tab{\ulocal} buf, b \in \nfst{\amessage}\\
\zeddashline
\tab{\uinit} buf = \nempty;~b = \nempty\\
\zeddashline
\uasm \\
\abox{1}  \t1   \msg{1}{a} 
\\
\zeddashline 
\ugar\\
\cbox{1}  \t1  \msg{1}{as}
 \\
 ~\\
 \forall t \in \Nat:
 \\ 
\cbox{2}  \t1  \neven{t} \to \ti{as}{t} = \nempty\\
~ \\
\cbox{3}  \t1   \nodd{t} \to \ti{as}{t} = b
 \\
~\\
\cbox{4}  \t1    \ti{r}{t} = \nempty \to
  b' = b ~\wedge~ buf' = newbuf \\
~\\
\cbox{5}  \t1    \ti{r}{t} \neq \nempty ~\wedge~ buf = \nempty \to  
            b' = \ti{a}{t} ~\wedge~ buf' = \nempty \\
          ~\\
\cbox{6}  \t1    \ti{r}{t} \neq \nempty ~\wedge~ buf \neq \nempty \to  
  b' = \nft{newbuf} ~\wedge~ buf' = \nrt{newbuf} \\
 ~\\
  \nwhere newbuf = \nif \ti{a}{t} = \nempty~ \nthen buf~ \nelse PrAdd(buf,~\nft{\ti{a}{t}})~ \nfi
\end{spec}

 The auxiliary function $PrAdd$ specifies the buffer update according to the priorities of the messages. A lower value of the identifier means a higher priority.
 
 \begin{schema}{\spc{PrAdd}}
\nfst{\amessage} \times \amessage \to \nfst{\amessage}\\
\ST
\auxbox{1}  \t1 PrAdd(\nempty, a) = \angles{a}\\
\auxbox{2}  \t1 PrAdd(\angles{x}\nconc y, a) =\\
\t3 \nif id(a) < id (x)\\
\t3 \nthen \angles{a} \nconc \angles{x} \nconc y\\
\t3 \nelse \angles{x} \nconc PrAdd(y,~a)\\
\t3 \nfi
\end{schema}

\newpage 
\section{Specification of a CAN component}

\begin{figure}[ht!]
    \centering
    \begin{spec}{\spc{CANArch}}{gb}
        \includegraphics[scale=0.7]{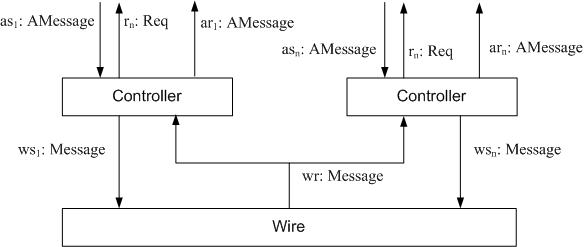}
    \end{spec}
    \caption{Logical architecture of a CAN component}
    \label{fig:CAN}
\end{figure}

Figure \ref{fig:CAN} presents the specification $CANArch$, which describes a logical architecture of a CAN protocol component. 
Each system node will be coordinated using the corresponding $Controller$ component, where the communication between controllers will go through the $Wire$ component.

The $Wire$ component has two assumptions on the input streams:
\begin{itemize}
    \item 
    all streams $ws_i$, $1 \le i \le n$ (CAN messages sent by $Controller$ components, where $n$ is the number of controllers, i.e., the number of CAN nodes in the system) must have at most one message per each time interval;
    \item
    at each time interval, if one of the streams $ws_i$, $1 \le i \le n$ is nonempty and carries an element of type $\Nat$ then all other streams $ws_j$, $1 \le j \le n$, $j \ neq i$ must be either empty or carry an element of type $\Nat$;
    \item
    at each time interval, if one of the streams $ws_i$, $1 \le i \le n$ is nonempty and carries an element of type $Data$ then all other streams $ws_j$, $1 \le j \le n$, $j \ neq i$ must be either empty or carry an element of type $Data$;
\end{itemize}

\begin{spec}{\spc{Wire}}{td}
\InOut{ws_1 ,..., ws_n : \cmessage}{wr: \cmessage}
\uasm \\
\abox{1}  \t1 \forall i \in [1..n]: \msg{1}{ws_i}
 \\
 ~\\
\abox{2}  \t1 \forall t \in \Nat:\\
  \t3 \exists i \in [1..n].~\ti{ws_i}{t} \neq \nempty \to \\
  \t5 
    (\nft{\ti{ws_i}{t}} \in \Nat \to 
 \forall j \in [1..n].~(\ti{ws_j}{t} = \nempty \vee \nft{\ti{ws_i}{t}} \in \Nat)) \\ 
 \t5 \wedge \\
 \t5 (\nft{\ti{ws_i}{t}} \in Data \to 
 \forall j \in [1..n].~(\ti{ws_j}{t} = \nempty \vee 
   \nft{\ti{ws_i}{t}} \in Data))
\\
\zeddashline 
\ugar\\
\cbox{1}  \t1 \msg{1}{wr}\\
\cbox{2}  \t1 \ti{wr}{0} = \nempty\\ 
\cbox{3}  \t1  \forall t \in \Nat:  \ti{wr}{t+1} = Broadcast(currentData)\\
\nwhere \\
currentData = CollectElements(n,\ti{ws_1}{t}, \dots, \ti{ws_n}{t})
\end{spec}

\begin{schema}{\spc{Broadcast}}
\nfst{\cmessage} \to  r: \nfst{\cmessage} \\
\ST
\auxbox{1}  \t1 Broascast(\nempty) = \nempty\\
\auxbox{2}  \t1 Broascast(\angles{x} \nconc y)\\
        \t3  \nif x \in \Nat\\
       \t3 \nthen \angles{MinNatList(x, y)}\\
       \t3 \nelse \angles{x}\\
       \t3 \nfi\\  
\end{schema}

\begin{figure}[ht!]
    \centering
    \begin{spec}{\spc{ControllerArch}}{gb} 
    \includegraphics[scale = 0.8]{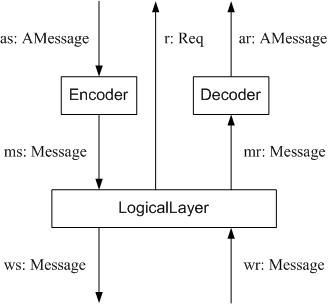}
    \end{spec}
    \caption{Logical architecture of a Controller component}
    \label{fig:controller}
\end{figure}

A $Controller$ component is also composite, the specification of its logical architecture is presented in 
Figure \ref{fig:controller}. 
$Controller$ consists of three sub-components:
\begin{itemize}
    \item $Encoder$ that converts the automotive application messages into CAN messages,
    \item $Decoder$ that ensures the reverse transformation, where CAN messages are decoded into the automotive application messages,
    \item $LogicalLayer$ that ensures that CAN bus behaves correctly.
\end{itemize}

The $Encoder$ component assumes that its input stream of type $\amessage$ can have at most one message per time interval. 
As soon as this component receives a message, it forwards its identifier to the logical level it the same time interval and sends the actual data part in the next time interval.
If we specify this behaviour simply by
\[
\begin{array}{ll} 
\forall t \in \Nat: & 
 \\
\cbox{(1)}   & \t1 \ti{as}{t} = \nempty \to \ti{ms}{t} = \nempty\\
\cbox{(2)}  & \t1 \ti{as}{t} \neq \nempty \to  \ti{ms}{t} = \angles{id(\ti{as}{t})} ~\wedge~ \ti{ms}{t+1} = \angles{data(\ti{as}{t})}
\end{array}
\]
We will have many contradictions. Thus, assume that $\ti{as}{t} \neq \nempty$ and $\ti{as}{t+1} = \nempty$. 
From (1) we can conclude that $\ti{ms}{t+1} = \nempty$. 
However, from (2) it follows that  $\ti{ms}{t+1} = \angles{data(\ti{as}{t})}$.
Also, in the case $\ti{as}{t} \neq \nempty$ and $\ti{as}{t+1} \neq \nempty$, we would have $\ti{ms}{t+1} = \angles{data(\ti{as}{t})}$ because $\ti{as}{t} \neq \nempty$, and at the same time 
$\ti{ms}{t+1} = \angles{id(\ti{as}{t+1})}$ because $\ti{as}{t+1} \neq \nempty$.
Thus, we have to use a state variable to ensure the correct modelling. Let us call this variable $e$. A simple Boolean type will be enough to specify the correct behaviour: the $\ntrue$ value will denote the state of active encoding process, where the $\nfalse$ value (which will be also the initial value for $e$) would mean that no encoding is currently performed.

\begin{spec}{\spc{Encoder}}{td}
\InOut{as: \amessage}{ms: \cmessage}
\tab{\ulocal} e \in \Bool;  
\zeddashline
\tab{\uinit} e = \nfalse \\
\zeddashline
\uasm \\
\abox{1}  \t1   \msg{1}{as} 
\\
\zeddashline 
\ugar \\
\cbox{1}  \t1   \msg{1}{ms}
 \\
~\\
 \forall t \in \Nat:
 \\
\cbox{2}  \t1  (e = \nfalse ~\wedge~ \ti{as}{t} = \nempty) \to \\
\t3
(\ti{ms}{t} = \nempty ~\wedge~ e' = \nfalse)\\
~\\
\cbox{3}  \t1  (e = \nfalse ~\wedge~ \ti{as}{t} \neq \nempty) \to \\
\t3
(\ti{ms}{t} = \angles{id(\ti{as}{t})}  ~\wedge~ \ti{ms}{t+1} = \angles{data(\ti{as}{t})}  ~\wedge~ e' = \ntrue)\\
~\\
\cbox{4}  \t1   e = \ntrue \to (\ti{ms}{t} = \angles{data(\ti{as}{t-1})} ~\wedge~ e' = \nfalse)\\ 
\end{spec}

The aim of the $Decoder$ component is to build an output message of type $\amessage$ out of two consequently received input messages, where the first input message must be of type $\Nat$ and the second input message must be of type $Data$.
This property is specifies as the following predicate:  

\begin{schema}{\spc{MsgCANFormat}}
s \in \ntist{\cmessage}
\ST  
\forall t \in \Nat:\\
\abox{1}  \t1 \ti{s}{t} \neq \nempty \wedge \nft{\ti{s}{t}} \in \Nat ~\to\\
      \t5 \ti{s}{t+1} \neq \nempty \wedge \nft{\ti{s}{t}} \in Data)
\\
\abox{2}  \t1 \ti{s}{t} \neq \nempty \wedge \nft{\ti{s}{t}} \in Data ~\to\\
      \t5 t > 0 \wedge \ti{s}{t-1} \neq \nempty \wedge \nft{\ti{s}{t-1}} \in \Nat )  
\end{schema}

Thus, the $Decoder$ component assumes that at each time interval it can receive at most one message, and if the message is non-empty and of type $\Nat$, the next time interval will of the input stream will contain data.
We will use a local variable $d$ of type $\Bool$ to denote that the decoding process is in progress: the $\ntrue$ value will denote the state of active decoding process, where the $\nfalse$ value (which will be also the initial value for $d$) would mean that no decoding is currently performed.

\begin{spec}{\spc{Decoder}}{td}
\InOut{mr: \cmessage}{ar: \amessage}
\tab{\ulocal} d \in \Bool\\%
\zeddashline
\tab{\uinit} d = \nfalse \\%
\zeddashline
\uasm \\
\abox{1}  \t1  \msg{1}{mr}\\
~\\
\abox{1}  \t1  MsgCANFormat(mr)
\\
\zeddashline
\ugar \\
\cbox{1}  \t1   \msg{1}{ar}\\
~\\
\forall t \in \Nat:\\
\cbox{2}  \t1   \ti{mr}{t} = \nempty \to (\ti{ar}{t} = \nempty ~\wedge~ d' = \nfalse)\\
\cbox{3}  \t1   \ti{mr}{t} \neq \nempty  ~\wedge~ d = \nfalse \to\\
        \t5 (\ti{ar}{t} = \nempty ~wedge~ d' = \ntrue)\\
\cbox{4}  \t1   \ti{mr}{t} \neq \nempty  ~\wedge~ d = \ntrue \to\\
        \t5 (\ti{ar}{t} = \angles{msg(\nft{\ti{mr}{t-1}},~\nft{\ti{mr}{t}} )} ~\wedge~ d' = \nfalse)\\
\\
\end{spec}

The $LogicalLayer$ component assumes that both its input stream of type $\cmessage$ can have at most one message per time interval and fulfil the property $MsgFormat$. 
All three its output streams also should have at most one message per time interval, where the $mr$-stream  that goes to the $Decoder$ component should in addition fulfil the property $MsgFormat$.

\begin{spec}{\spc{LogicalLayer}}{td}
\InOut{ms, wr: \cmessage}{mr, ws: \cmessage;~ r: Req}
\tab{\ulocal} lid \in \Nat\\
\zeddashline
\tab{\uinit}  lid = 0 \\
\zeddashline
\uasm \\
\abox{1}  \t1  \msg{1}{ms}\\
\abox{2}  \t1  \msg{1}{mr}\\
\abox{3}  \t1   MsgFormat(ms)\\
\abox{4}  \t1   MsgFormat(mr)\\ 
\\
\zeddashline 
\ugar\\
\cbox{1}  \t1  \msg{1}{ws}\\
\cbox{2}  \t1  \msg{1}{r}\\
\cbox{3}  \t1  \msg{1}{mr}\\
\cbox{4}  \t1  MsgFormat(mr)\\
\cbox{5}  \t1  \forall t \in \Nat:~\ti{mr}{t} = \ti{wr}{t}\\
~\\
  \textsf{tiTable~LLTable}
\end{spec}

\begin{tabular}{l |l| l || l| l || l ||l}
\multicolumn{7}{l}{\textsf{tiTable~LLTable}: $\forall t \in \nat$}\\
\hline
\hline
& $ms$ & $wr$ & $ws$ & $r$ & $lid'$ & Assumption\\
\hline
 1 &  \nempty & $y$	& \nempty  & \nempty & $lid$ &\\
 2 &  $x$ & $y$	& \nempty  & \nempty & \nft{x} & $x \neq \nempty$, $\nft{x} \in \Nat$\\
 3 &  $x$ & \nempty	& \nempty  & \nempty & $lid$ & $x \neq \nempty$, $\nft{x} \in Data$\\
 4 &  $x$ & $y$	& $x$  & \angles{req} & $lid$ 
 & $x \neq \nempty$, $\nft{x} \in Data$, $y \neq \nempty$, $\nft{x} = lid$\\
 5 &  $x$ & $y$	& \nempty  & \nempty & $lid$ 
 & $x \neq \nempty$, $\nft{x} \in Data$, $y \neq \nempty$, $\nft{x} \neq lid$\\
\hline
\end{tabular}

~\\
~\\

\noindent
\textbf{Remark:}\\
The 3rd line of the table \textsf{LLTable} will never be used by the specification 
\textit{LogicalLayer} because of the assumptions and the properties of the specification
\textit{Wire}.

\section{Related work}
\label{sec:related}

\subsection{CAN}

There have been very few formal approaches targeting analysis of CAN protocol. 
A formal method for analysis of automotive systems (also CAN-based) was discussed in \cite{hamann2006formal}.
A frame packing algorithms for automotive applications was introduced in \cite{saket2006frame}. 

Van Osch and Smolka proposed a finite-state method for analysis of the CAN bus protocol. 
Saha and Roy presented a formal
specification of the time triggered version of CAN Protocol, see \cite{saha2007finite}.

\subsection{\focust}

\focust approaches presented in \cite{hffm_spichkova,spichkova2013design,spichkova2013we}
aims to apply the engineering psychology achievements to the design of formal methods, focusing on the specification phase of a system development process.
Its core ideas originated from the analysis of the \Focus\ framework and also led to 
an extended version of the framework, \focust.  

Another approach based on \focust, allows analysis of component dependencies    \cite{spichkova2014formalisation}. This was later extended to framework for formal analysis of dependencies among services \cite{spichkova2014towards}.

Model-based analysis of temporal properties using \focust was presented in \cite{spichkova2017model}. 
The authors also demonstrate how to implement on \focust basis time-triggered and event-based view on systems with temporal properties.

Spatio-temporal models for formal analysis and property-based testing were presented in 
\cite{alzahrani2016spatio,alzahrani2017temporal} by Alzahrani et al.
The authors aimed to to apply property-based testing on \focust and TLA models  with temporal properties. 

Zamansky et. al. \cite{zamansky2016formal,zamansky2016formal} reviewing some recent large-scale industrial projects in which formal methods (including \focust) have been successfully applied. The authors also covered some aspects of teaching formal methods for software engineering, including \focust, cf. \cite{spichkova2016teaching,simic2016enhancing}.

\section{Conclusions}
\label{sec:conclusions}

This paper presents a formal specification of the Controller Area Network (CAN) protocol using \focust\ framework.   
We formally describe core components of the protocol, which provides a basis for further formal analysis using the Isabelle/HOL theorem prover \cite{npw} using the \emph{Focus on Isabelle} methodology \cite{spichkova,spichkova2013stream,spichkova2008focus}.

\newpage
\bibliographystyle{abbrv}

\end{document}